\documentclass[aip,prl,reprint]{revtex4-1}

\usepackage{graphicx}
\usepackage{subfigure}
\usepackage{amsmath}

\begin{document}

\title{Optically controllable photonic structures with zero absorption}

\author{Chris O'Brien}
\email[]{cobrien@physics.tamu.edu}
\affiliation{Department of Physics and Astronomy and Institute for Quantum Studies, Texas A\&M University, College Station, TX 77843-4242}
\author{Olga Kocharovskaya}
\affiliation{Department of Physics and Astronomy and Institute for Quantum Studies, Texas A\&M University, College Station, TX 77843-4242}

\begin{abstract}
We show the possibility to periodically modulate refractive index in atomic medium in space or/and time while simultaneously keeping vanishing absorption/gain. Such modulation is based on periodic resonant enhancement of refractive index, controlled by an external optical field, and opens a way to produce coherently controllable photonic structures. We suggest possible implementation of the proposed scheme in rare-earth doped crystals with excited state absorption. 
\end{abstract}
\maketitle
%
Index of refraction $n = Re\sqrt{1 + \chi}$ is a real number defined by the susceptibility i.e., the linear response of a medium to electromagnetic radiation which is strongly enhanced near atomic resonances. However, the enhancement of refractive index near atomic resonances is accompanied by enhancement of absorption. Such that, when the maximal contribution from the atomic resonance to the refractive index is reached, the contribution to the absorption is on the same order which prevents the usage of obtained refractive index in  transmission experiments. 

There have been several proposals on how to resonantly enhance refractive index while at the same time eliminating resonant absorption. \cite{Scully, Rathe, Lukin, Dowling} One approach taken is based on interference effects in multilevel atomic systems driven by coherent resonant or off-resonant fields at the adjacent transitions where absorption could be canceled by inversionless or Raman gain.\cite{Fleischhauer} Another suggestion is based on compensation of absorption with resonant gain at an inverted transition shifted with respect to an absorbing transition on the scale of its linewidth.\cite{Yavuz} Such a situation could be realized either in a mixture of two two-level atomic species, or in a single atomic species possessing simultaneously both noninverted and inverted transitions with slightly shifted frequencies, such as in rare-earth doped crystals with excited state absorption.\cite{OBrien} Proof of principle experiments were done for schemes with both resonant and off-resonant driving in hot Rb vapors in which enhancement of refractive index $n \sim 10^{-4}$ was achieved under negligible absorption.\cite{Zibrov, Proite} An enhancement up to the value $n \sim 10^{-2}$ is expected with an increase of a density to $N = 6 \cdot 10^{16}$cm$^{-3}$. The further increase of refractive index in room-temperature gases with increase of the density is not feasible due collisional broadening becoming the dominant contribution to the linewidth. Much higher resonant additions to the background index are anticipated in transition element doped crystals due to the essentially higher density of the ions which does not in general result in proportional line broadening.\cite{OBrien, Crenshaw, Rebane}      

Coherent control of the magnitude of refractive index is attractive for creating a new type of  photonic crystal or distributed Bragg reflector, with optically controllable properties. However, it turns out to be difficult to maintain transparency for varying values of the refractive index. Also, in many of the previously suggested schemes transparency at a particular frequency is typically accompanied by gain in a neighboring frequency range leading to instability. In this work we suggest a scheme which is free of these drawbacks. It allows for optical modulation of refractive index while keeping transparency of the medium and suppressing the development of instabilities associated with the presence of gain.

%
To illustrate the idea we first consider the interaction of a probe field with a medium of three level atoms in a ladder configuration such that the probe field interacts with both transitions as illustrated in the inset of fig. 1. The transition frequencies $\omega_{21}$ and $\omega_{32}$ are close to each other so that the probe field with frequency $\omega _p$ interacts simultaneously with both transitions and for a weak probe Rabi frequency $\Omega _p << \gamma _{21}, \gamma _{32}$ or $\Omega _p << \gamma _{31}$ the susceptibility is defined as the sum of the susceptibilities of two two-level transitions: 
\begin{equation}
\chi = \frac{3N\lambda ^3 }{8\pi ^2} [ \frac{\gamma ^{\text{rad}} _{21} (\rho _{1} - \rho _2)}{\delta _{21} -i\gamma  _{21}} + \frac{\gamma ^{\text{rad}} _{32} (\rho _{2} - \rho _{3}) }{\delta _{32} -i\gamma  _{32} }].
\end{equation}
Here N is the atomic density, the detunings are defined as $\delta_{21} = \omega _{21} - \omega _p$ and $\delta_{32} = \omega _{32} - \omega _p$, $\lambda$ is the probe field wavelength in the medium, $\gamma ^{rad} _{ij}$ is the radiative decay rate for the i to j transition, $\gamma _{ij}$ is the total decoherence rate, and $\rho_{i}$ is the population in the i$^{\text{th}}$ energy level. We assume that the amplitudes of both transitions are matched but of opposite sign:
\begin{equation}
\gamma _{21} ^{\text{rad}}(\rho_{1} - \rho_{2}) = -\gamma _{32} ^{\text{rad}} (\rho_{2} - \rho _{3}).
\end{equation}
which means that one of the two transitions is inverted. Let it be transition 2-1, i.e. $\rho_{2} - \rho _{1} > 0$. We also assume the widths of the transitions are equal $\gamma _{21}  = \gamma _{32}$ and the probe field is tuned to two photon resonance i.e., $\omega _p = \omega _{31}/2$. 
Thus for arbitrary position of level 2 the blue detuning of the probe field from one of two two-level transitions is equal to the red detuning from another one i.e., $\delta _{32} = -\delta _{21} = \delta$, leading to the remarkable property that gain at one transition and absorption at another one cancel each other while the real part of susceptibility is doubled. Thus the susceptibility is purely real and takes the form:
\begin{equation}
\chi = \frac{3N\lambda ^3 \gamma ^{\text{rad}} _{21} (\rho_1 - \rho _2)}{8\pi ^2} \frac{2\delta}{\delta ^2 + \gamma_{21}  ^2}.
\label{shifteq}
\end{equation}
%
\begin{figure}
\begin{center}
\resizebox*{8cm}{!}{\includegraphics{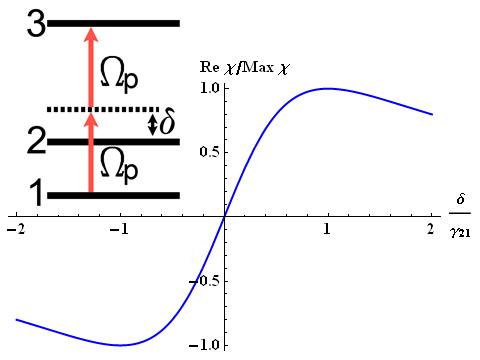}}
\label{fig1}%
\caption{Plot of the real part of the susceptibility as a function of the level shift $\delta$. Note the imaginary part is identically zero. The inset gives the energy level diagram for the corresponding three level scheme.}%
\end{center}
\end{figure}
%
\begin{figure}
\begin{center}
\resizebox*{8cm}{!}{\includegraphics{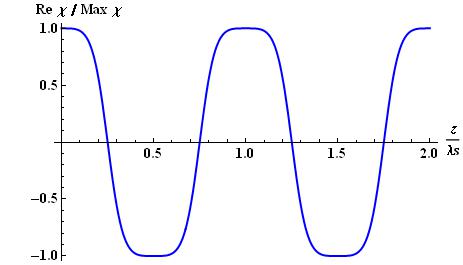}}
\label{fig2}%
\caption{Real part of the susceptibility plotted as a function of position along the optical axis.}%
\end{center}
\end{figure}
\begin{figure}
\begin{center}
\resizebox*{8cm}{!}{\includegraphics{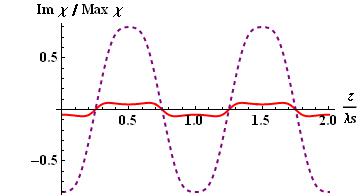}}
\label{fig3}%
\caption{Imaginary part of the susceptibility for a probe field detuned from resonance by $\gamma _{21}/20$ (solid) and for a detuning of $\gamma _{21}$ (dashed)  plotted as a function of position along the optical axis.}%
\end{center}%
\end{figure}%
It means that the probe field neither experiences absorption nor gain independently of level 2's energy i.e., for arbitrary values of $\delta$. At the same time the resonant susceptibility varies from the minimum to the maximum value as $\delta$ is shifted from $-\gamma $ to $\gamma$ as shown in fig. 1. If the energy of the intermediate level is modulated in space along the direction of propagation of the probe field, 
the refractive index is also modulated. Such spatial modulation can be produced along the optical axis via the ac-Stark shift. A control laser field $E _s cos(\omega _s t)$ applied at the 0-2 transition adjacent to the 1-2 transition and far detuned from this transition $\Delta _s = \omega_{s}-\omega_{20} >> \gamma _{20}$ would result in a splitting of the intermediate state 2 into two ac-Stark sublevels shifted in frequency by $-|\Omega _s|^2/\Delta _s$ and $\Delta _s+|\Omega _s|^2/\Delta _s$, respectively, where $\Omega _s$ is the associated Rabi frequency. The probe field is far out of resonance with the transitions from the second Stark sublevel from both level 1 and level 3 and, therefore its interaction with these transitions is negligible while the first Stark sublevel is slightly shifted from the original level 2 and strongly interacts with the probe field. In other words, the susceptibility at each transition (2-1 or 2-3), which in general consists of two terms associated with the one-photon and two-photon resonances is reduced to the one-photon contribution and has the same form as eq. \ref{shifteq}, just with shifted transition frequencies.

If the control field represents itself as a standing wave such that the Rabi frequency is a function of position inside the medium, $\Omega _s (z) = \Omega _s cos(k_sz)$, then the ac-Stark shift of level 2 is given by:
\begin{equation}
\Delta E =  -\frac{\hbar |\Omega _s|^2}{2\Delta _s} - \frac{\hbar |\Omega _s|^2}{2\Delta _s}cos(2k _s z).
\end{equation}
Thus it consists of a constant shift, $|\Omega _s|^2/2\Delta _s$, and a sinusoidal modulation, $(|\Omega _s|^2/2\Delta _s)cos(2k_sz)$. If the difference between the atomic transition frequencies $\omega_{32} - \omega_{21}$ is chosen to be equal to $-|\Omega _s|^2/\Delta _s$ then the susceptibility is described by eq. 3 with $\delta = (|\Omega _s|^2/2\Delta _s)cos(4\pi z/\lambda_s)$ (where $\lambda_s$
is the wavelength of the control field in the medium). Hence the refractive index will be modulated symmetrically with respect to its background value as shown in fig. 2.  The spacial period $\lambda _s/2$ is defined by the wavelength, while the modulation depth $-|\Omega _s|^2/\Delta _s$ is defined by the Rabi frequency of the modulating field $\Omega _s$. To provide the maximum amplitude of refractive index modulation the Rabi frequency of the control field should meet the condition $\Omega_s^2 = 2\gamma \Delta_s$.

With a strong index variation a transparent, for a particular frequency, 1-D photonic crystal can be created with properties that are optically controlled. Similarly a 3-D photonic structure can be produced by application of 3 orthogonal modulating control fields. Even for small index variations the medium will behave as a distributed Bragg reflector if the spacial period coincides with half of the probe wavelength in the medium i.e., $\lambda _s \simeq \lambda_p$. 
Since the medium remains transparent, many periods of spacial RI structures can be used as needed to achieve perfect reflection.
For collinear propagation of the probe and control field the Bragg condition implies equal wavelengths and hence equal frequencies of the control and probe field. It can be met in the case of orthogonal polarizations of these fields when transitions 1-2 and 0-2 satisfy proper selection rules (such as, for example, transitions from Zeeman sublevels of the ground state $|n=1,m=1>$ and $|n=1,m=-1>$ coupled to an excited state $|n=2, m=0>$ by right and left circular polarized fields accordingly) and corresponds to a detuning $\Delta_s$ equal to the frequency separation between Zeeman sublevels (which should be taken to be much larger than the optical linewidth $\gamma _{21}$). Another option is to use a modulating control field with higher frequency then the probe field $\omega_s - \omega_p >> \gamma_{21}$ in a slightly non-collinear geometry so that the projection of the wave vector of the control field would be equal to $k_p$.

When the probe field is detuned from two-photon resonance with 1-3 transition it will experience either gain or absorption. The question arises if such gain may result in the building up of a spontaneously amplified field emptying the inverted transition and limiting the propagation length of the probe field in the medium with periodic refractive index. Fortunately, this is not the case. Indeed, since the position of the intermediate level is periodically modulated in space, then a detuned probe field experiences periodically interchanging regions of gain and absorption suppressing the development of such an instability as can be seen in fig. 3. 

The simple model of a ladder system previously discussed assumed the existence of two transitions possessing equal linewidths, equal products of transition strength and population difference and  nearly degenerate (on the scale of the linewidth) frequencies. It is difficult if not impossible to meet these conditions in a real atomic system. However, it is possible to construct an effective ladder system whose upper transition has controllable parameters which could be optically tuned to satisfy these conditions.

It can be accomplished by adding to the original simple ladder system along with the modulating control field $E_s$ coupled to an adjacent transition 0-2 (as discussed above) a second control field $E_c$ coupling the excited state 3 to an additional unpopulated level 4 as shown in fig. 4. This second far-detuned control field ($\Delta_c>> \gamma^{rad} _{32}, \Omega_c$ where $\Delta _c = \omega _{43} - \omega _c$ and $\Omega _c$ is the control field Rabi frequency) is chosen to satisfy approximately the two-photon resonance condition: $\omega_c-\omega_p=\omega_{42}$, forming together with the probe field a far-detuned lambda scheme.  A strong far-detuned field results in an ac-Stark splitting of level 3 and the response to the probe field consists of two terms representing one-photon (upper Stark sublevel) and two-photon (lower Stark sublevel) contributions in the same way as previously discussed. But now it is the two-photon contribution which plays a dominant role due to the two-photon resonance condition.\cite{OBrien, Anisimov}

As a result, the total five level system under the formulated above conditions is reduced to an effective three-level ladder system with the lower transition 1-2'' and the upper transition 2''-3'. Its susceptibility takes the form:
\begin{widetext}
\begin{equation}%
\chi _{\text{res}} = \frac{3N \lambda ^3 }{8 \pi ^2} \left \{ \frac{\gamma _{21}^{\text{rad}}p/(2-p)}{\delta_p +\frac{\Omega_s^2}{2\Delta_s}+\delta - i\gamma _{21}} + \frac{\xi \gamma _{32} ^{\text{rad}} /[(2 - p)(1+2\xi)]}{\delta_p  - \omega_{32} + \omega_{21}-\frac{\Omega_s^2}{2\Delta_s} -\delta + \Delta_c(1+\xi-\xi^2) -i[\gamma_{42}(1-\xi)+\gamma_{32}\xi]} \right \}.
\label{analytic}
\end{equation}
\end{widetext}
For simplicity we assume level 3 is empty and introduce the pumping parameter $p=(\rho _{2} - \rho _{1})/\rho _{2}$, a control field parameter $\xi = |\Omega _c|^2/\Delta _c ^2$, and the one photon detuning $\delta_p = \omega_{21} - \omega_p$.
Now the parameters of the effective upper (2''-3') and lower (1-2'') transitions defined by the control fields can easily be matched.

We chose $\Omega _s = \sqrt{2\gamma _{21} \Delta_s}$ to provide the maximum range of refractive index modulation.
Matching the linewidth of 3'-2'' transition to that of 2''-1 defines the control field parameter $\xi $: 
\begin{equation}
\xi = \frac{\gamma _{21}-\gamma _{42}}{\gamma _{32} - \gamma _{42}}.
\label{Omegac}
\end{equation} 
It implies a larger linewidth of the upper 2''-3' transition as compared to the lower transition 1-2'': $\gamma _{32}>\gamma _{21}$ and  relatively slow decay of the coherence at the 4-2'' transition: $\gamma _{42}<\gamma _{32},\gamma _{21}$.
Matching the amplitudes defines the pump parameter as:
\begin{equation}
p= \frac{\gamma _{32}^{\text{rad}}}{\gamma_{21}^{\text{rad}}}\frac{\xi}{1+2\xi}.
\label{p}
\end{equation} 
We take the probe field to be resonant with the dressed transition 2''-1 such that $\delta _p = -\Omega_s^2/2\Delta_s$.
Then matching the frequencies of the transitions defines the required detuning of the control field $\Delta_c$:
\begin{equation}
\Delta _c = \frac{\omega _{32} - \omega _{21} + 2\gamma _{21}}{1+\xi - \xi^2}.
\label{Delta}
\end{equation} 
This implies that $\Delta_c$ will be on the same order as $\omega_{32}-\omega_{21}$. Since $|\Omega_c| = \sqrt{\xi} \Delta_c$, $\Omega_c$ and $\Delta_c \approx \omega _{32} - \omega _{21}$, it is important to have 1-2 and 2-3 transitions with close frequencies in order to reduce the required control field intensity.
Under the above conditions the susceptibility (5) takes the same form as eq. \ref{shifteq}. Thus, it becomes possible to realize resonant modulation of refractive index with zero absorption/gain in the realistic system. 
%
\begin{figure}
\begin{center}
\resizebox*{5cm}{!}{\includegraphics{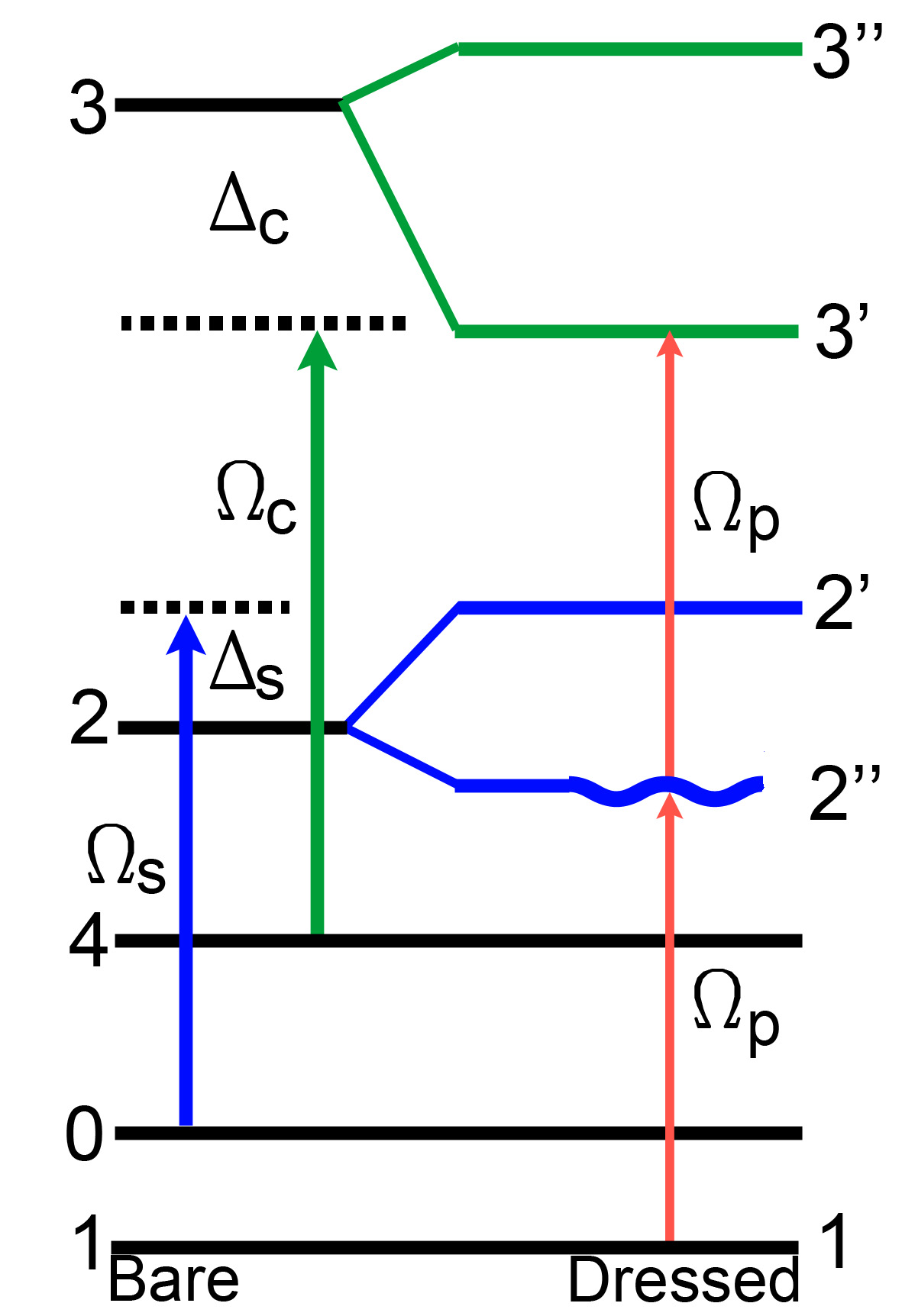}}
\label{fig4}%
\caption{Energy level diagram for the 5-level system coupled with two control fields $\Omega_s$ and $\Omega_c$ leading in ac Stark splitting of levels 2 and 3  and resulting in an effective ladder system 1-2''-3' in the dressed state basis.}
\end{center}%
\end{figure}%
%
\begin{figure}
\begin{center}
\resizebox*{8cm}{!}{\includegraphics{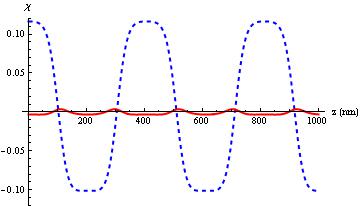}}
\label{fig5}%
\caption{Real (dashed) and imaginary (solid) part of the susceptibility as a function of distance along the optical axis, for implementation of a  optically controlled distributed Bragg grating in Er$^{3+}$:YAG with the numbers listed in the paper.}
\end{center}%
\end{figure}%

As an example we consider Er$^{3+}$:YAG (n$_{\text{bg}}$ = 1.82) where the $ ^4 I _{9/2}$ to $^4 I _{15/2}$ ($\gamma _{21} ^{\text{rad}} = 45$Hz) transition at 813.2nm (transition 2-1 in fig. 4) has a closely matched excited state absorbtion transition (transition 2-3 in fig. 4) from $^4 I _{9/2}$ to $^4 G _{9/2}$ ($\gamma _{32} ^{\text{rad}} = 15$Hz) with $\omega _{32} - \omega _{21} = 20$GHz.\cite{Sarder, Gruber} Coherent driving of the transition between the next Stark level of the ground state and  $^4 I _{9/2}$  level (transition 2-0 in fig.4) can be used for modulation of level 2 position, while coherent driving of $^4 I _{15/2}$ and $^4 G _{9/2}$ can be used for matching of the parameters of the upper and lower transitions in the effective ladder system.  
Assuming $N = 1.4 \cdot 10^{21}$cm$^{-3}$, $\gamma _{32} = 0.8$GHz, $\gamma _{21} = 0.3$GHz, and $\gamma_{42} = 0.2$GHz and choosing pump parameter, p = 0.035, and the following parameters of the driving fields: $\Omega _s = 2.45$GHz, $\Delta _s = 10$GHz, $\Omega _c = 7.449$GHz, and $\Delta _c = 17.893$GHz, we obtain 3.3\% refractive index modulation with respect to background value ($\Delta \chi ' = 0.22$) with a periodically modulated practically vanishing absorption (max $|\chi''| < 0.0033$) as shown in fig. 5. This result follows from the numerical analysis of the 5 level system driven with two coherent fields, and is well approximated by the analytical formula in eq. \ref{analytic}. For an interaction length of 100$\mu$m there will be 245 periods of modulation which leads to a distributed Bragg reflector with a reflection coefficient of R = 0.99998. Since there will be absorption/gain as the probe field is moved away from the transparent frequency, the distributed Bragg reflector will have a very narrow bandwidth of 0.6GHz.     

In conclusion, we proposed a method to produce periodic modulation of the refractive index while keeping zero net absorption/gain. The method is based on modulation of the populated intermediate energy level of an effective three level system with matched transition properties in space by an external strong control field via the ac-Stark effect. Possible implementation of this technique in Er$^{3+}$:YAG is suggested, where a 3\% modulation of refractive index with vanishing absorption is possible. The proposed method may find useful applications for the creation of optically controlled distributed Bragg reflectors and photonic crystals.

This research was supported by NSF grant No. 0855688.

\end{document}